\documentclass[prb,twocolumn,showpacs,preprintnumbers,amsmath,amssymb]{revtex4}

\usepackage{graphicx,rotating}
\usepackage{dcolumn}
\usepackage{bm}
\usepackage{epsfig}
\usepackage{longtable}

\begin{document}

\title{Hole $sp^3$-character and delocalization in (Ga,Mn)As 
revised with pSIC and MLWF approaches - 
newly found spin-unpolarized gap states of $s$-type below 1$\%$ of Mn}

\author{Karolina~Z.~Milowska\footnote{karolina.milowska@gmail.com} }
\affiliation{%
Institute of Theoretical Physics, Faculty of Physics,
University of Warsaw, ul. Ho\.za 69, 00-681 Warszawa, Poland \\
Photonics and Optoelectronics Group, Faculty of Physics,        
Ludwig-Maximilians-University Munich,
Amalienstrasse 54, 80799 Munich, Germany
}%
\author{Ma\l{}gorzata~Wierzbowska\footnote{malgorzata.wierzbowska@fuw.edu.pl} }
\affiliation{%
Institute of Theoretical Physics, Faculty of Physics,
University of Warsaw, ul. Ho\.za 69, 00-681 Warszawa, Poland
}%

\begin{abstract}
The dilute magnetic semiconductor (Ga,Mn)As is ferromagnetic in accordance
with the p-d Zener model. Hole density function (HDF) localization
has been previously studied by means of the density functional theory (DFT)
and non-standard DFT methods; however not for dopings near 1$\%$.
We have revised (Ga,Mn)As using the DFT with the pseudopotential
self-interaction correction (pSIC) and maximally-localized Wannier functions
(MLWFs), which show the $sp^3$ character of a HDF.
Nature of HDF is extended - for low dopings and the pSIC, 70$\%$ of the HDF
is located within the inter-impurities region, and contribution of the 3d-Mn
states is 3-5$\%$ for 1-3$\%$ of Mn with the pSIC, and 11$\%$ with the DFT.
We found that for dopings below 1$\%$, the spin-unpolarized $s$-type impurity
states segregate from the conduction band to the energy gap - in contrast to
earlier publications. This implies that donor co-doped dilute samples would be
both insulating and nonmagnetic.
\end{abstract}


\maketitle

\section{Introduction}
\label{sec:int}

The dilute magnetic semiconductors (DMS) possess combined semiconducting and
magnetic properties, hence their potential application in spintronics
have been extensively investigated over past 
30-years \cite{RRG,Furdyna,Ohno,Dietl-Nature,Sato-RMP}. 
A prototype DMS used to explain a mechanism of the ferromagnetic order in
this class of magnetic materials is (Ga,Mn)As.
It is a half-metal, with a metallic phase in majority-spin band, 
and with a spin-polarized hole introduced to the host GaAs by a substitution of Ga with Mn.
The nature of this hole rules magnetic and transport properties. Therefore,
deep understanding of a type and localization of empty states near the Fermi
level and within the energy gap is a subject of hot debate.   

In the pioneering work, Dietl et al. \cite{Dietl-Science,Dietl-Tc} assumed 
the extended-hole scenario, with a HDF delocalized within 
the valence band, and applied the p-d Zener model to successfully explain 
high Curie temperature, reaching 110~K, in GaAs with 5$\%$ of Mn.
Since then, many papers based on this model have been written and very
intensive ab-initio studies on the (Ga,Mn)As system have been performed 
(see a review  \cite{RMP-TD}).
However, perfect consensus about this system is not achieved, and very
low dopings were not studied in detail by means of first-principles methods, 
while there are many experimental papers considering this type of 
systems \cite{oiwa, hamaya, mac, prb, reid}.

It is also stated in a few works that some impurity states at low dopings
 appear within the energy gap. Their character is believed to be of valence band
origin and spin-polarized and the $d$-type. 
In contrast to those investigations, 
we have found that the gap states close to 1$\%$ doping originate 
from the conduction band and their character is spin-unpolarized 
and the $s$-type.

In our studies, we have employed the following two additional methods beyond standard 
implementations of DFT (like GGA), which have not been previously applied 
to the hole localization problem at very low concentrations:
\begin{enumerate}
	\item pSIC scheme \cite{pSIC,implement},
which corrects the electronic self-interaction problem for all atomic shells,
giving a more balanced treatment than just the $d$-shell correlations
included in the DFT+U approach
  \item MLWFs analysis \cite{MLWF}
\end{enumerate}
These two methods complete the HDF localization description in 
the real-space manifold of localized functions centred at atoms or bonds.
We have focused on two Mn concentrations of 1$\%$ and 3$\%$, showing
that, at some co-doping with donors, the samples may exhibit different magnetic and 
conducting properties.

Next in the scope of this work, we have listed earlier studies on
the HDF localization in (Ga,Mn)As in section 2,
given details of our calculations in section 3, presented
our results  in section 4, and drawn conclusions in section 5.

\section{Brief summary of earlier studies} 

The earlier DFT and DFT+U calculations for (Ga,Mn)As at dopings of 6$\%$ and 12$\%$ 
already have shown that the highest contribution to the HDF originates
from the As-neighbours of Mn \cite{Park}, 
and this result was confirmed in the system with 3$\%$ of Mn \cite{Dublin}.
Further studies for concentrations 3-12$\%$ showed that the HDF extends
beyond the second As neighbours of the impurity \cite{Sandratskii}.
First application of the pSIC method to this system for dopings of 6-12$\%$ 
presented the contribution of the 4$p$-Mn states to the top of 
the valence band \cite{Alessio}.  

Other self-interaction corrected scheme, namely SIC-LSD, 
has been applied to  (Ga,Mn)As at 3-25$\%$ of Mn,
and only the majority-spin $d$-states have been treated with this method, 
beyond standard DFT \cite{Szotek}.
That work, however, focuses on critical temperature and a comparison of (Ga,Mn)As
to Mn doped GaP and GaN, showing the total density of states (DOS) 
and the $d$-projected DOS of Mn only.
The standard DFT calculations for lower Mn doping of 2$\%$ have been also 
performed \cite{Hill}.   

None of the  publications mention the gap states, 
since the considered dopings were too high to find them.
Those works did not focus on the chemical character of a hole, 
as it was considered in previous study \cite{Stroppa} for a different system: Mn in Ge.
Very detailed investigations of chemical character of 
the 3$d$-Mn states in work \cite{Zunger-old} do not consider explicitly 
a contribution of the 4$p$-Mn states to the HDF.
We found these states dominating over the t$_{2g}$-Mn 
contribution when the pSIC scheme was applied.

Nevertheless, all listed above publications agree about 
the HDF delocalization scenario, which leads to justification
of the p-d Zener model for T$_c$ in (Ga,Mn)As
\cite{Dietl-Tc,Jungwirth,Molenkamp}. 
Our results presented here support this model too.

Recently, a contradicting hypothesis, with the Fermi level localized
within the 3$d$-Mn impurity states, is discussed by the experimental group
\cite{Samarth}, and followed by a theoretical work \cite{Zunger-new}
with similar conclusions to earlier publication \cite{Kikoin}. These works
examine the double-exchange mechanism as possible in (Ga,Mn)As.  

As for the gap states at lower dopings, some publications imply that the
origin of these states is from the valence band and their spin-polarization  
is of the $d$-type \cite{Molenkamp,Zunger-new}. 
We have found that these states are spin-unpolarized and of the $s$-type,
and they are formed by a separation from the conduction band.

\section{Theoretical Details}

We have started performing the calculations within the density functional 
theory framework \cite{DFT}, employing the {\sc Quantum ESPRESSO} 
code \cite{qe}, with the pseudopotentials (PPs) and the plane-wave basis. 
For the exchange-correlation functional, we have chosen 
the generalized gradient approximation (GGA) by means of 
the Perdew-Burk-Erzenhof parametrization \cite{PBE}. We have used the ultrasoft
pseudopotentials \cite{USPP} with the semicore 3$s$- and 3$p$-shell 
of Mn and the 3$d$-shell of Ga included in the valence bands, and with
nonlinear core correction for the Ga and the As PPs. The energy cutoffs 
of 35 Ry and 350 Ry were set for the plane-waves and the density, 
respectively. 

The calculations were done for two Mn concentrations: 
1$\%$ with one Mn atom in the 216-atoms cell and 3$\%$ with a single impurity
in the 64-atoms cell. Dense k-point grids of (6,6,6) and (12,12,12) points 
generated according to the Monkhorst and Pack scheme \cite{grid}
were used, for these dopings respectively. 
For a quadrature over the Brillouin zone,  
the metallic-smearing technique \cite{smearing} close to the Fermi surface was
used with the Gaussian broadening of 0.01 Ry. The experimental lattice
constant of 5.65~\AA$\;$ was fixed for all calculations,
and the atomic positions in the cells were optimized. The largest 
relaxation, found close to the Mn impurity, was smaller than 0.6$\%$ 
of the Ga-As bond length. 

The pSIC scheme \cite{pSIC,implement} results were compared to the GGA results.
The self-interaction corrections have been applied to all atomic shells of all
atoms in the calculated cells and to both spins.

To imagine a character of the hole, we performed the detailed analysis
of the projected density of states onto the atomic shells. 
The hole occupation numbers $n_h^{PDOS}$ were obtained
from a quadrature of the projected DOS, $N(\varepsilon)$, in a range
from the Fermi level $\varepsilon_F$ to the energy at which the DOS vanish
first time for the unoccupied states: 
\begin{equation}
  n_h^{PDOS} \; = \; \int_{\varepsilon_F}^{\varepsilon_{N(\varepsilon)=0}}
  \; N(\varepsilon) \; d\varepsilon.
\end{equation}

The shape of HDF and the degree of localization in the real space have been examined using
the maximally-localized Wannier functions (MLWFs) \cite{MLWF} for which
calculations are made with the aid of the Wannier code \cite{wannier}.
The electronic localization degree can be estimated from the MLWFs spreads,
$\Omega_n$, defined as \cite{MLWF}:
\begin{equation}
 \Omega_n \; = \; [\langle r^2 \rangle_n - {\bf \bar{r}}^2_n ],
\end{equation}
where
${\bf \bar{r}}_n^2=\langle 0n|{\bf r}|0n\rangle^2 = \langle{\bf r}_n\rangle^2$
and $\langle r^2 \rangle_n = \langle 0n| r^2| 0n \rangle$, with $|0n\rangle$
being the Wannier function with number $n$ and centered in the original cell
with the direct-lattice vector ${\bf R}=0$, and ${\bf r}$ is
the real-space position operator. The sum of above defined quantities
$\Omega = \sum_n \Omega_n$ is minimized in the MLWFs-finding
procedure \cite{MLWF}.

To get a closer insight into the Mn-As bonds, we calculated the MLWFs on the
GGA and the pSIC Bloch-functions obtained for the 64-atoms cell.
For minimization of the total spread, we have chosen the 133- and
128-band space in the spin up and down, respectively.
Bottom of the energy window was set within the gap between the
localized 3$d$-Ga derived bands and the delocalized $sp$-bands of (Ga,Mn)As.
Top of the energy window was fixed just above the 133-rd band
counted for the spin up from the bottom of the energy window.
From this band-manifold, we obtained 133 and 128 MLWFs in the
spin up and down, respectively. 

In analogy to the hole occupation numbers obtained from the DOS, $n_h^{PDOS}$,
we analyse contributions of the HDW from the MLWFs, $n_h^{MLWF}$, defined via
the MLWFs occupations as $n_h^{MLWF}=1-f_n^{MLWF}$.
Concept of the Wannier occupations has been introduced recently for
proper symmetries of some systems, which need a few unoccupied states (usually
the anti-bonding counterparts to the valence states) in the manifold
of Bloch functions to be used for the MLWFs construction \cite{Arash}.
These occupation numbers $f_{nm}^{MLWF}$ are defined with the use of
two unitary transformation matrices $U_{pq}^{dis}$ and $U_{ij}$,
where the first acts in the disentangling procedure to obtain
the optimal subspace of Bloch-like functions possessing proper symmetries
and the second is obtained during the MLWFs optimization process.
Thus the occupation matrix is as follows:
\begin{equation}
f_{nm}^{MLWF} = \sum_{k\in BZ} \sum_p^{occ} \sum_{s,r}^{win}
 \; U_{rm}^k \; U_{pr}^{k \;dis} \;
U_{sn}^{k \; \ast} \; U_{ps}^{k \; dis \; \ast},
\end{equation}
where 'win' runs over all states in the outer window
(including some unoccupied states) and 'occ' runs over states up to
the Fermi level.
Since the off-diagonal occupations sum to zero in the whole system, we
use only the diagonal occupations ($f_n=f_{nn}$) in the definition of MLWFs
contributions to the HDF.

\section{Results}

The Mn impurity, replacing Ga in the GaAs host, offers two electrons
from the 4$s$-shell and five electrons from the 3$d$-shell instead of the
Ga configuration 4$s^2$3$d^{10}$4$p^1$. Since the valence of As is 5 and
that of Ga is 3, the substitution of Ga with Mn creates a hole in the
valence band, because five 3$d$ electrons of Mn almost do not take a part
in binding with the As neighbours. 

\begin{figure}
\includegraphics[scale=0.34,angle=0.0]{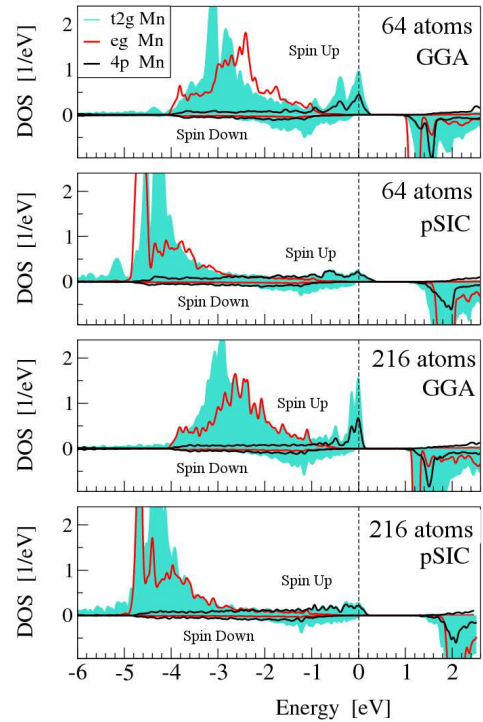}
\caption{(Color online)
Projected density of states (DOS) for t$_{2g}$, e$_g$ and 4$p$ states of Mn
in (Ga,Mn)As at doping levels of 3$\%$ (64-atom cell) and 1$\%$ (216-atom cell).
The Fermi level is marked by the vertical dashed-line.}
\label{DOS}
\end{figure}

In Figure~\ref{DOS}, the density of states (DOS), projected on the t$_{2g}$,
e$_g$ and 4$p$ Mn-orbitals, is presented. The Fermi level cuts through the
valence band top, for two concentrations and both theoretical methods
applied. The DOS centre of mass, for the t$_{2g}$ and e$_g$ states, 
moves in the pSIC scheme energetically downwards in comparison to the GGA.
Also, the number of states at the Fermi level decreases in the pSIC. 

These results are consistent with the previous pSIC \cite{Alessio} 
and SIC \cite{Szotek} and the LDA+U \cite{Park,Dublin,Sandratskii} calculations. 
The e$_g$-shell in spin up ($\uparrow$) is fully occupied,
and the t$_{2g}$ ($\uparrow$) states are almost completely filled,
with L\"owdin's occupation analysis \cite{lowdin} giving range of values 2.74-2.85.
The spin-down ($\downarrow$) states are quite empty for the Mn-3$d$ states,
with L\"owdin's occupations of 0.52-0.72 for the t$_{2g}$.
Total magnetisation in the cell is about 
4.0-4.2 $\mu_B$ for all methods and Mn concentrations. 
Absolute magnetisation is much higher, 4.86-5.36 $\mu_B$, 
also due to substantial polarization of neighbouring As atoms,
0.07 $\mu_B$, coupled antiferromagnetically to the impurity moment. 

\begin{figure}
\includegraphics[scale=0.43,angle=0.0]{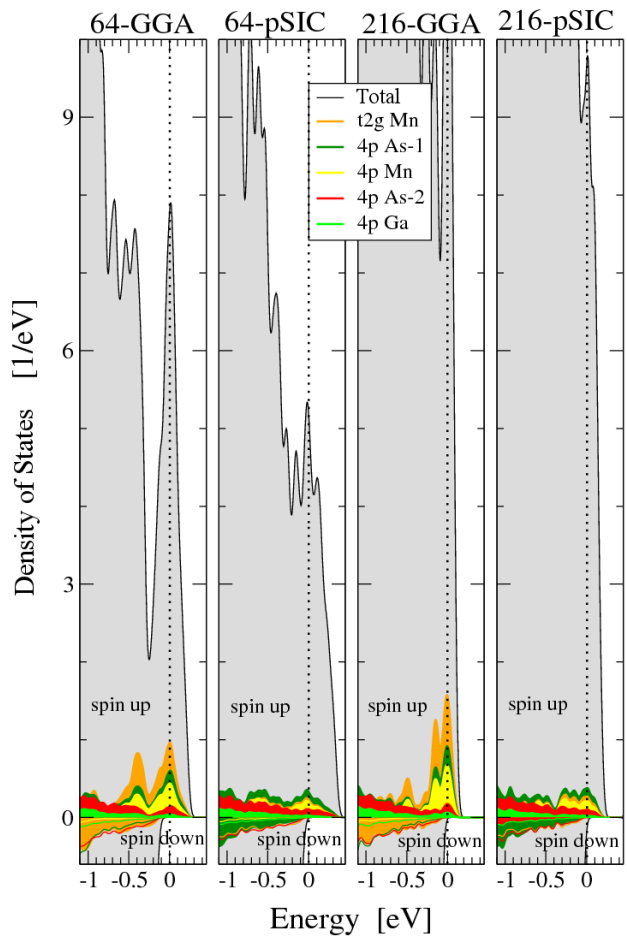}
\caption{(Color online)
Projected density of states (DOS) close to the Fermi level at dopings
of 3$\%$ and 1$\%$, obtained with the GGA and the pSIC.
The total DOS is marked by the grey colour.
As-1 and As-2 denote the first- and second-neighbour As-atoms along
the Mn-As-Ga-As chain in 110-direction, the Ga-atomic site is (1/2,1/2,0)
in units of the lattice constant and  Mn is at the origin.}
\label{fermi}
\end{figure}

\begin{table}
\caption{Contribution to the hole occupations $n_h^{PDOS}$ 
from the projected DOS [in $\%$], for two doping levels: 3$\%$ and 1$\%$,
obtained with the GGA and the pSIC.  
Wyckoff positions in the cells in units of the lattice constant are given
in parenthesis. The sum of contributions from Mn and its neighbours along
four tetragonal easy-axes are given in the last row.} 
\begin{tabular}{lcccc}
\hline
\hline \\[-0.2cm]
   & \multicolumn{2}{c}{64 atoms} & \multicolumn{2}{c}{216 atoms} \\
 States (Atomic Position) & GGA  & pSIC & GGA & pSIC \\[0.1cm]  
 \hline \\
 t$_{2g}$-Mn (0,0,0) &  10.75  & 4.65  &  11.10  &  3.03 \\
 4$p$-Mn (0,0,0)       &  4.97   & 4.05  &  4.78   &  2.34 \\ 
 4$p$-As (1/4,1/4,1/4) &  7.17   & 7.18  &  6.67   &  4.41  \\
 4$p$-Ga (1/2,1/2,0)   &  0.52   & 0.65  &  0.47   &  0.36 \\
 4$p$-As (3/4,3/4,1/4) &  2.05   & 2.89  &  1.34   &  1.32  \\
 Total from 4 easy axes &  54.68  & 51.58  & 49.80  &  29.73  \\[0.1cm] 
\hline
\hline
\end{tabular}
\label{occ}
\end{table}

Closer perspective at the Fermi-level region of the DOS projected onto 3$d$-
and 4$p$-Mn states, and onto 4$p$-states of neighbouring atoms
from  the Mn-As-Ga-As chain along the 110 axis, as well as the total DOS,
are presented in Figure~\ref{fermi}.
In both the DFT and the pSIC schemes, the impurity states are mainly
localized deeply below the Fermi level, 3-4 eV as seen in Figure~\ref{DOS},
therefore the hole states almost do not contain the Mn-component.
The hole states merge with the valence band for both concentrations 
of impurities, 1$\%$ and 3$\%$. This fact is better pronounced within 
the pSIC approach. 

For accuracy, Table~\ref{occ} collects the hole occupation numbers, 
$n_h^{PDOS}$, defined in the previous section. 
We conclude, that the HDF is composed of many states and
the contribution of the 4$p$-As states is the highest by means of the pSIC.
If we take into account the fact that there are four As nearest
neighbours of Mn, then it is obvious that the HDF is mainly located around
the impurity neighbours and not at the impurity.
The 4$p$-Mn donation to the HDF is almost as high as from the
3$d$-Mn states (pSIC) or about half of the Mn-3$d$ input (GGA).
Interestingly, the HDF extends beyond the second As-neighbours of Mn, and
only half or less of the hole occupation is summed over
the Mn atom and its twelve neighbours from the Mn-As-Ga-As chains along
four easy axes. Half of the HDF extends over the inter-impurity part of
the supercells at doping 3$\%$, or it is even 70$\%$ in case of 1$\%$ Mn 
calculated within the pSIC approach. Similar analysis has been performed in
the publications \cite{Sandratskii,Alessio} 
by means of the DFT, the DFT+U and the pSIC approaches.

\begin{figure*}
\begin{picture}(400,40)(0,0)
\put(3,10){$d_z^2$}
\put(70,10) {$d_{x^2-y^2}$}
\put(138,10){$d_{xy}$}
\put(198,10){$d_{xz}$}
\put(258,10){$d_{yz}$}
\put(318,10){$sp^3$ As-Mn}
\put(384,10){$sp^3$ As-Ga}
\end{picture}
\centerline{ \includegraphics[scale=0.10]{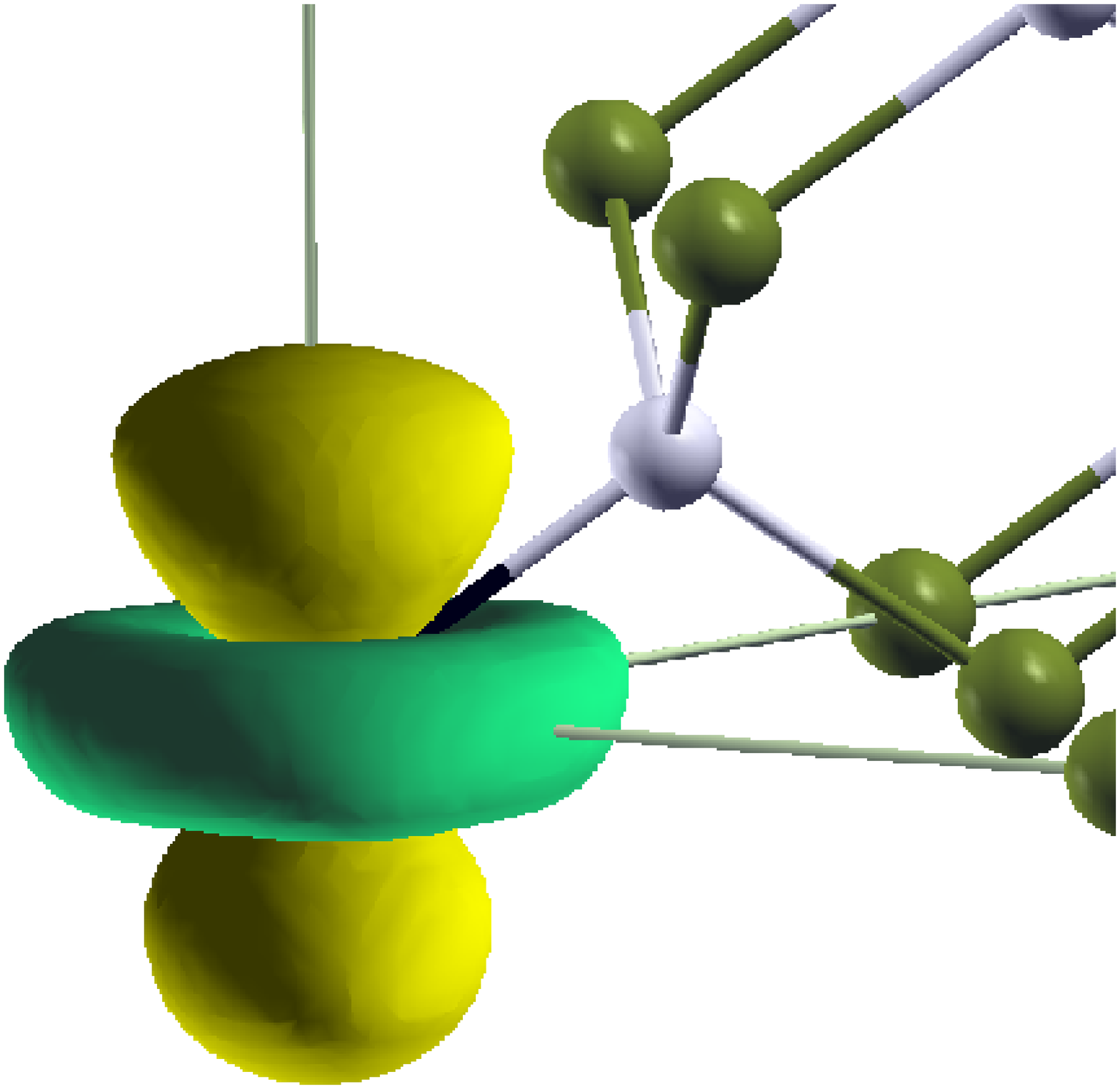}
\includegraphics[scale=0.10]{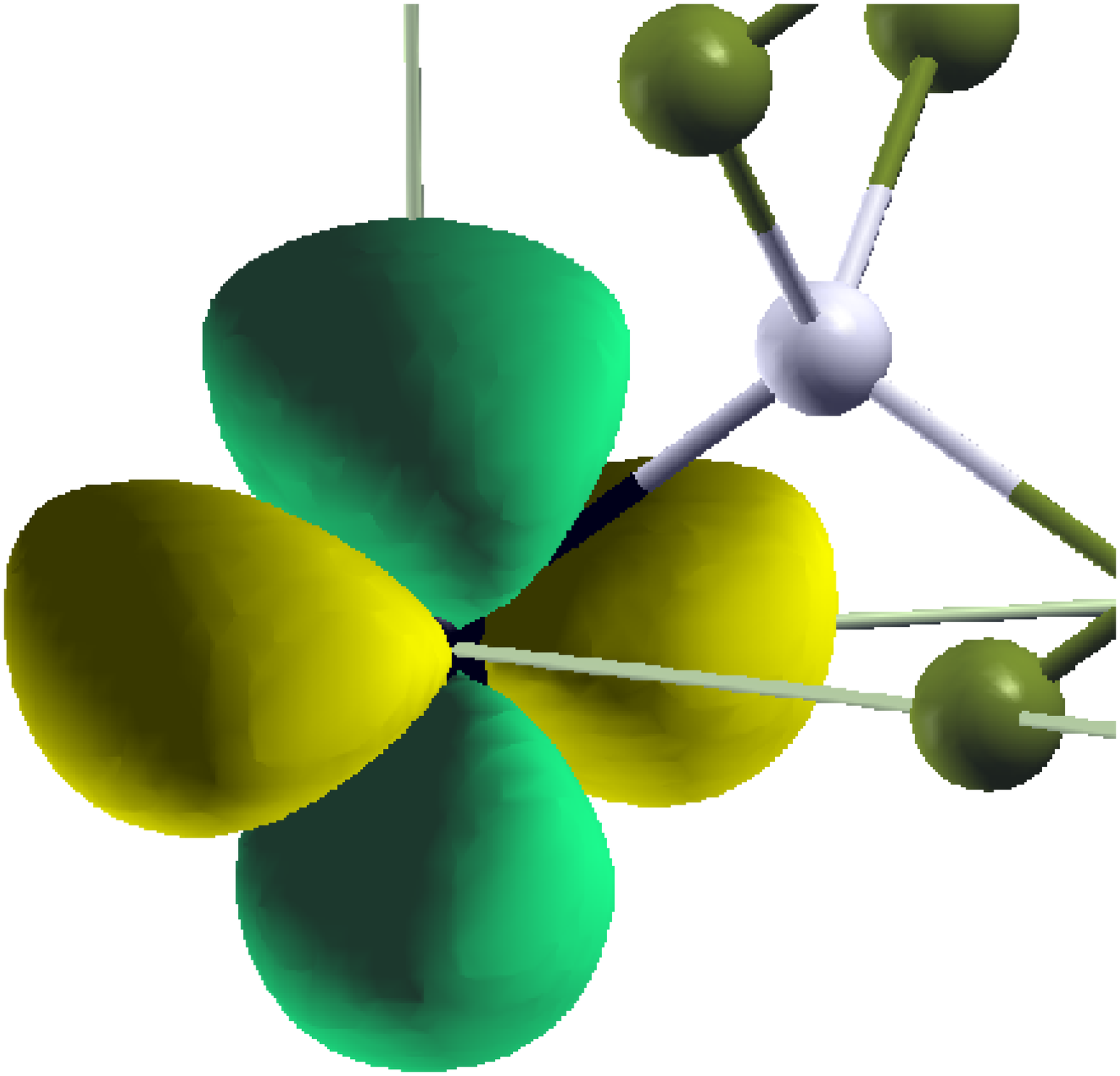}
\includegraphics[scale=0.10]{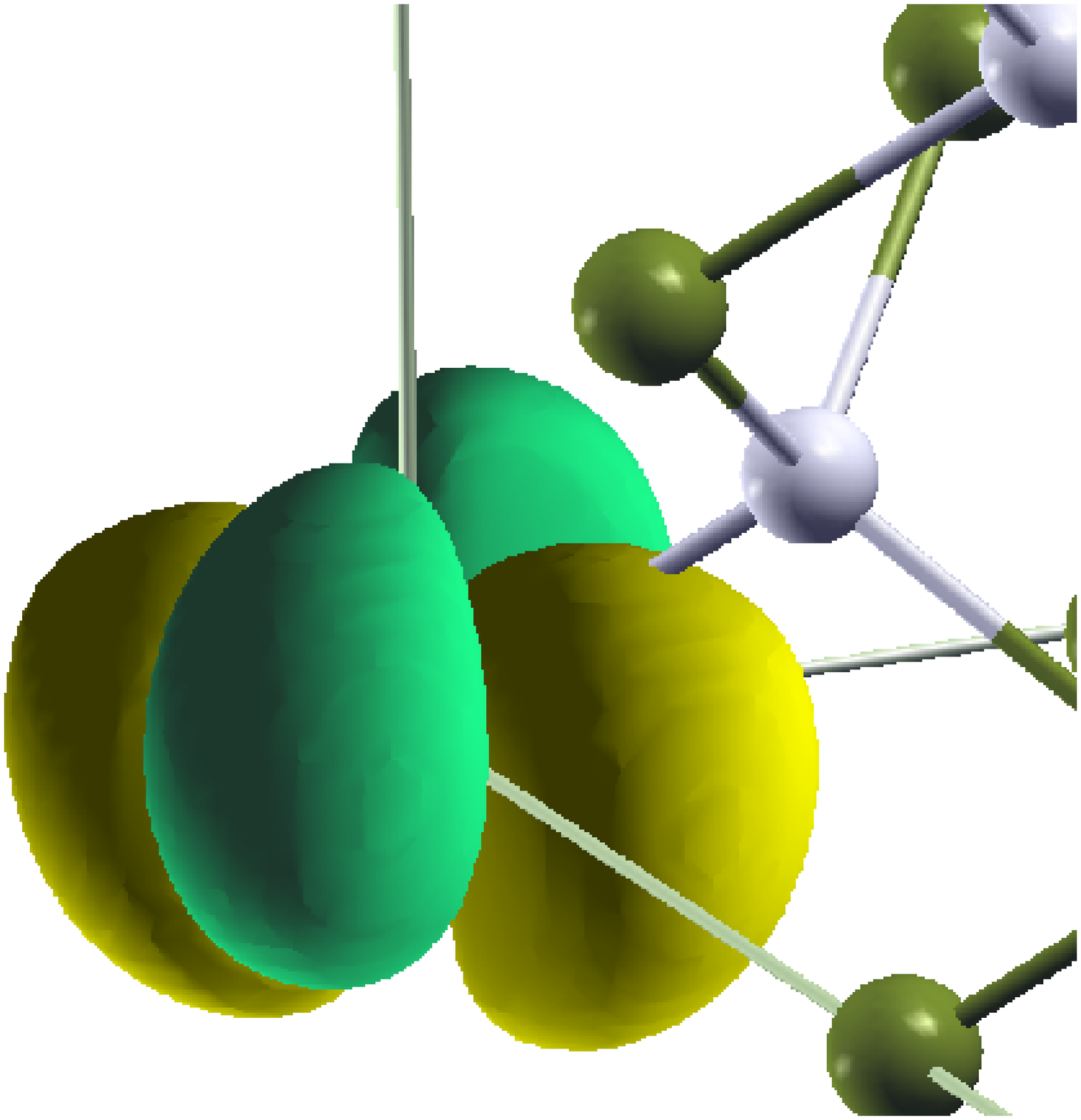}
\includegraphics[scale=0.10]{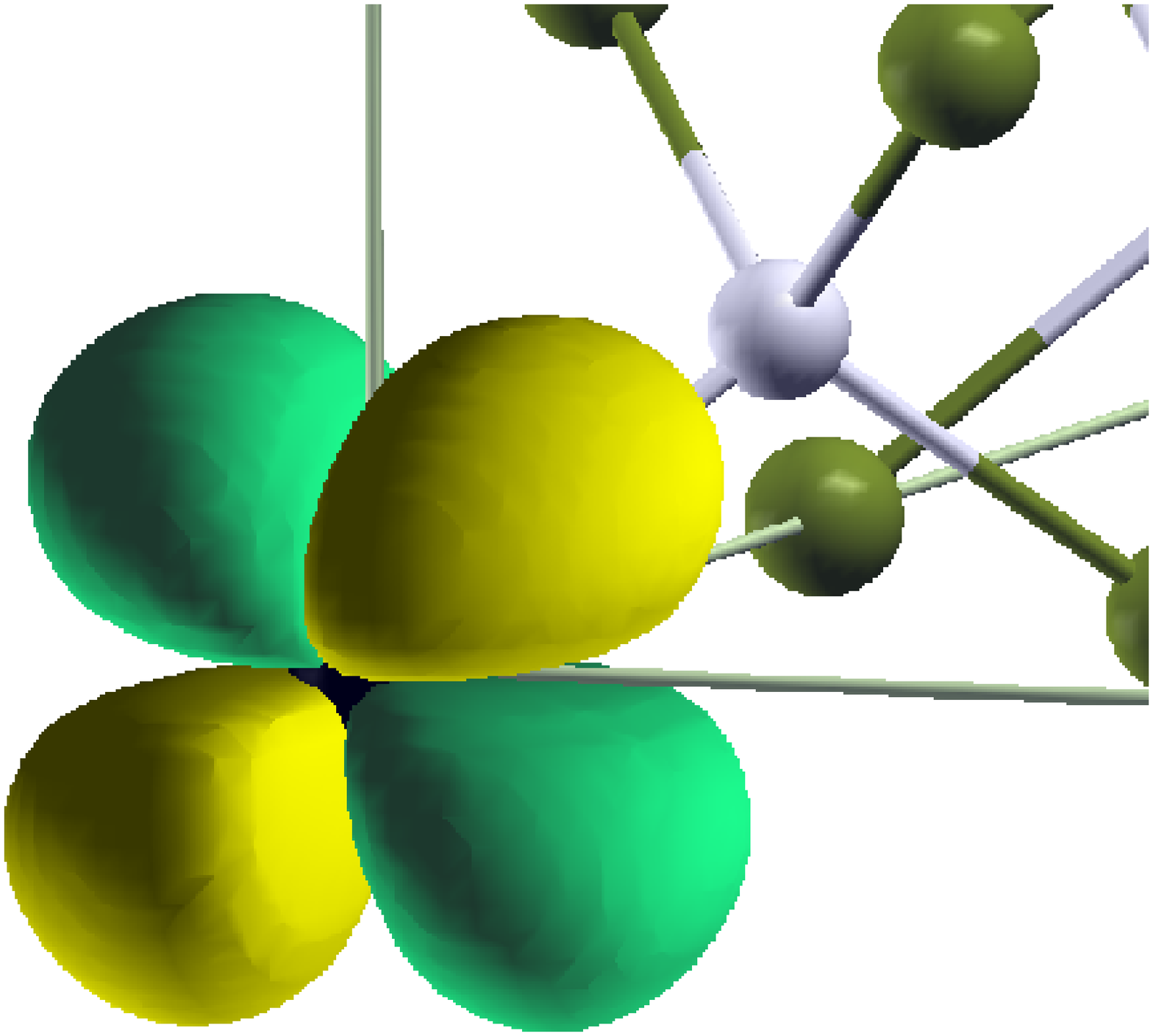}
\includegraphics[scale=0.10]{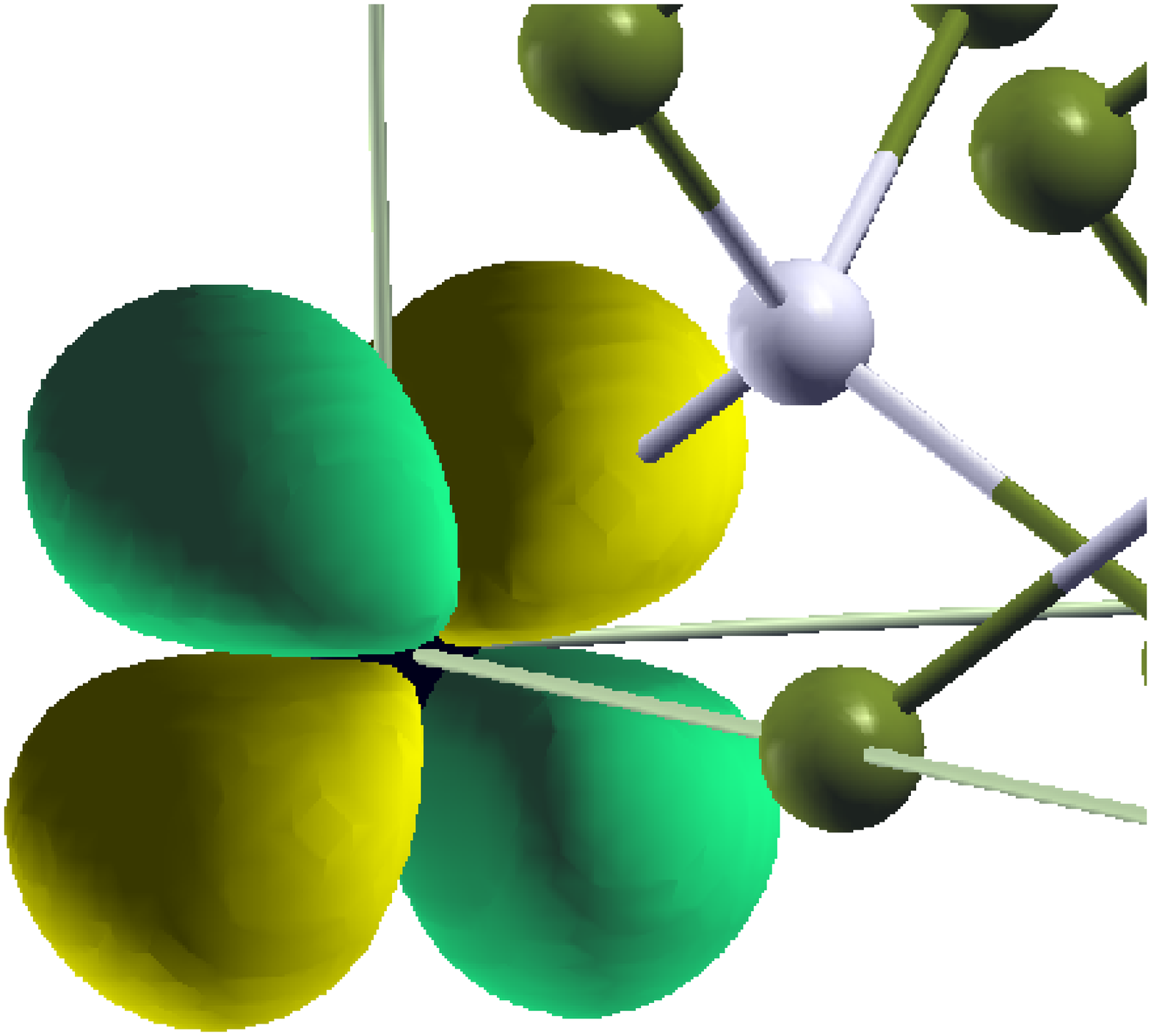}
\hspace{2mm}
\includegraphics[scale=0.10]{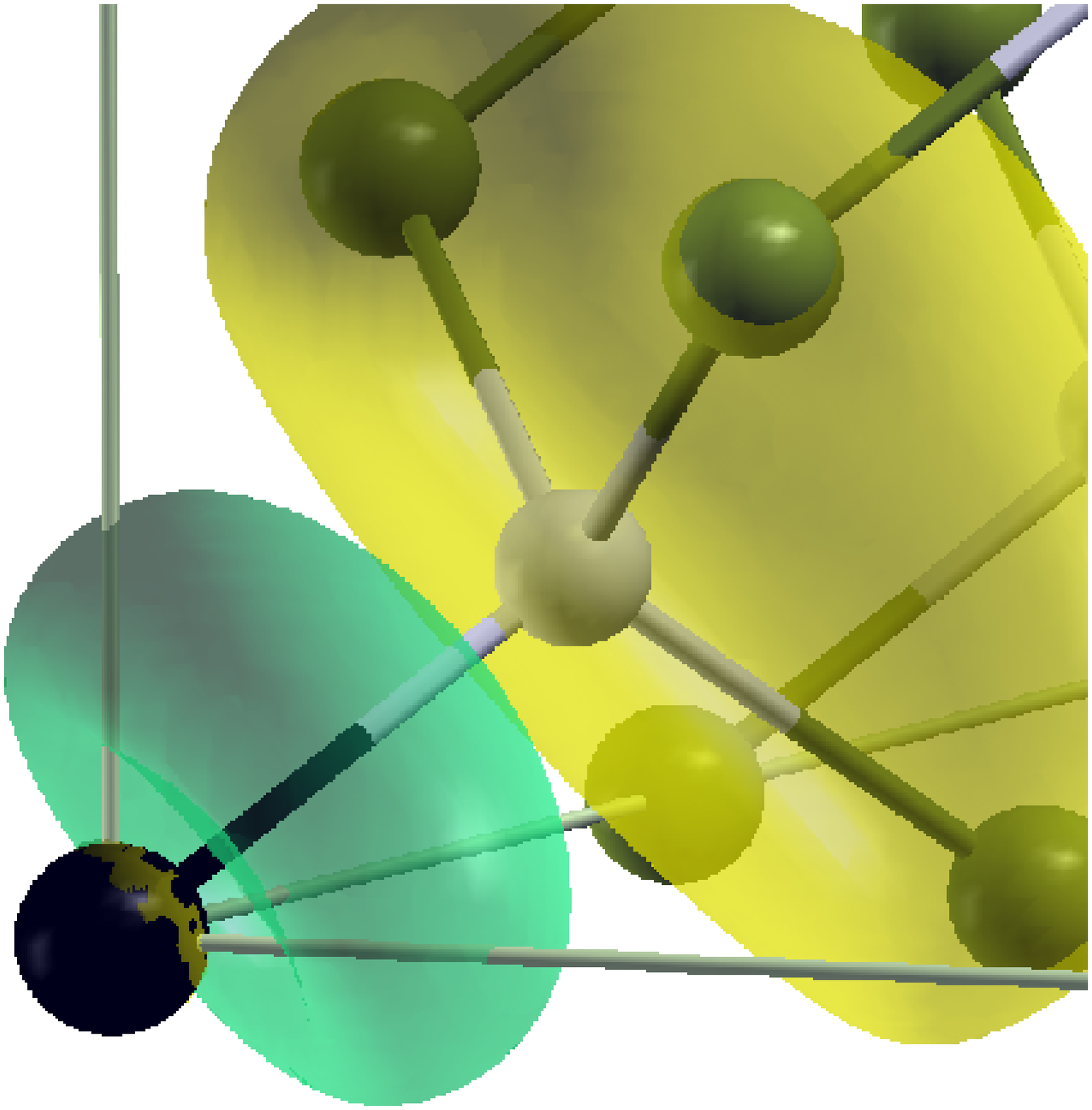}
\hspace{1mm}
\includegraphics[scale=0.10]{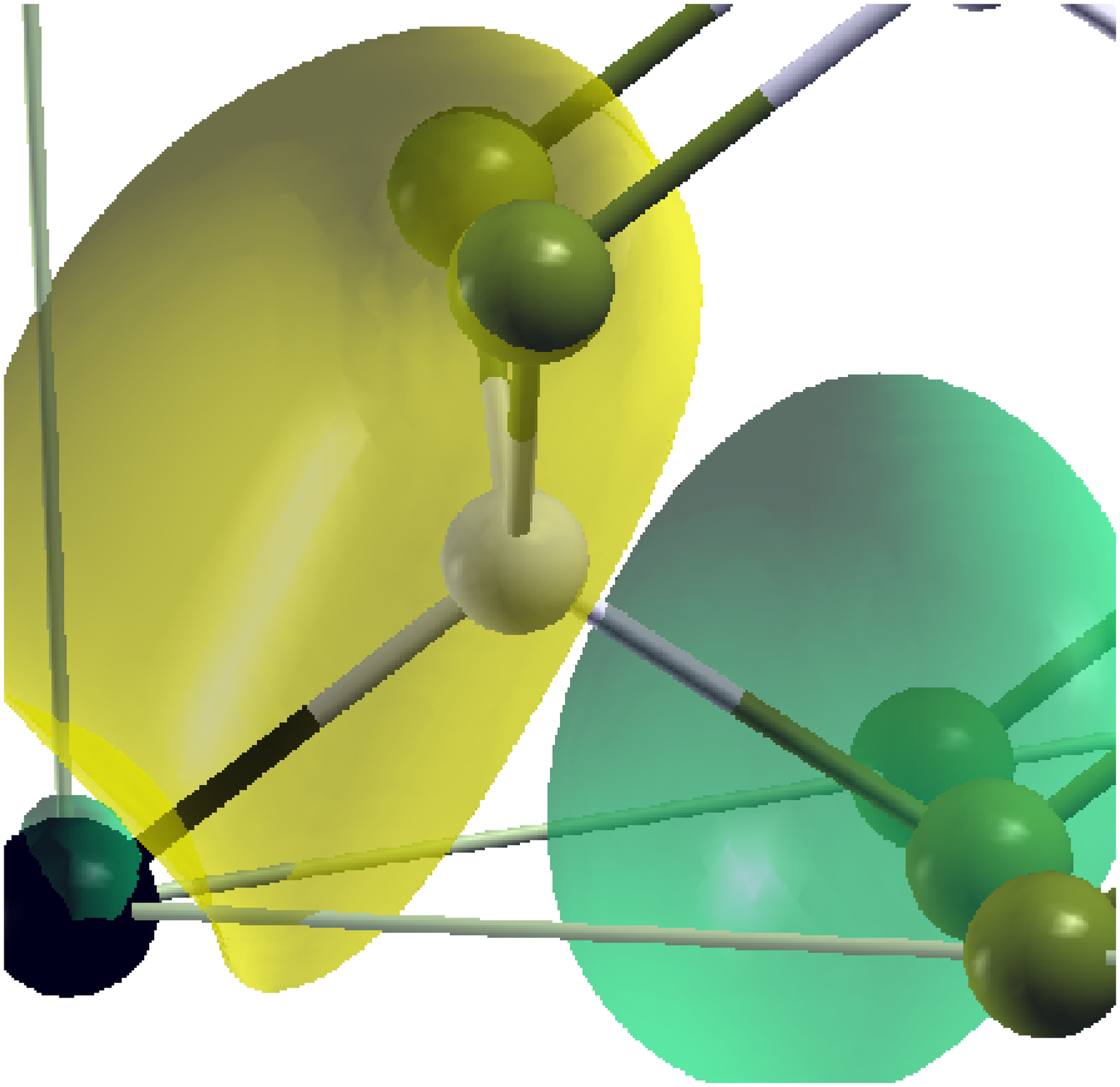}}
\caption{(Color online)
Maximally-localized Wannier functions, for the spin up, centered at Mn and
the neighbouring As atom, obtained for the 64-atoms cell with the GGA.
The $sp$-lobes along the As-Mn and As-Ga bonds were chosen from a manifold of
$sp^3$ hybridization. All functions were plotted with the same isovalues.
The Mn atoms are depicted in black colour, As in white and Ga in olive colour.
Plots were prepared with the xcrysden code \cite{xcrys}.}
\label{Wan}
\end{figure*}

Chemical bond of Ga with As-neighbours is built by 4$s$ and 4$p$ electrons.
For the Mn impurity, occupations of the 4$s$-states are about
0.34 ($\uparrow$) and 0.27 ($\downarrow$) for both the GGA and the pSIC,
and for the 4$p$-states the corresponding numbers are
0.70 ($\uparrow$) and 0.50 ($\downarrow$) (GGA) and
0.89 ($\uparrow$) and 0.60 ($\downarrow$) (pSIC), independently of
impurities concentrations.

The Mn-As and Ga-As bonds analysed with the  MLWFs appear as $sp^3$-type lobes 
centred closely to each of 32 As atoms in the cell, 
these centres are slightly on the back bonds.
The 3$d$-type functions centred at Mn rather do not take a part in bonding.
Similar $sp^3$-hybridization has been discussed by Stroppa et al. \cite{Stroppa}
for the Mn impurity in Ge.
Because of the chosen energy window, the Wanniers obtained by us contain the HDF. 
Plots for some MLWFs for the spin-up channel, obtained with the GGA,
are drawn in Figure~\ref{Wan}.

The spreads for some representative MLWFs for (Ga,Mn)As,
obtained from the GGA- and the pSIC Blochs, are collected in
Table~\ref{spreads} and compared to the MLWFs obtained for the isolated Mn
atom and pure GaAs.
It is clear, that characters of the $sp$-lobes centred on the Mn-As bonds are
very similar to those on the As-Ga bonds, except slightly smaller spreads
of the As-Mn MLWFs caused by a little shift away from Mn. The 3$d$-Mn {MLWFs}
are much more localized than the $sp^3$-functions, although, spreads of
the $d$-type functions are twice larger, for e$_g$, and three times larger, 
for t$_{2g}$, than for the corresponding functions of the isolated atom.

Plots also show that a contribution of the t$_{2g}$-symmetry functions
to the As-Mn bond is small, and even with a tendency to escape from
the bondline, as one can see from the asymmetry of plus- and minus-sign lobes
of the $d_{xy}$ MLWF ($d_{xz}$ and $d_{yz}$ have the same property).
Spreads of the $sp^3$ MLWFs on the As-Ga bonds in (Ga,Mn)As are larger
than these in pure GaAs, like Mn-substitution effect was blowing them.
It might be a signature of the HDF being localized close to the As atoms.
Finally, an effect of the pSIC shows better localization around atoms.
This causes decrease of atomic spreads, and increase of the
distance between lobes of the interatomic MLWFs, and decrease of covalence.
When bands are infinitely thin, and there is no k-points dispersion, then
an effect of the pSIC on the MLWF-spreads is vanishing (Mn atom).

To give a measure of the HDF localization with the MLWFs approach, 
we present, in Table~\ref{spreads}, the hole occupation numbers 
$n_h^{MLWF}$, defined in the previous section. 
The HDF is distributed over all bonds in the supercell. 
It's half-localization, in the 64-atoms supercell, 
extends to the second As-neighbours of the impurity (this volume contains 
13 atoms). Far away from the impurity 
(which is placed in the corner of a supercell),
at four central As atoms in the 64-atom supercell, 
there is about 4$\times$(0.37+3$\times$0.39)=6.16$\%$ of a HDF obtained with
the GGA method and $8.8\%$ by means of the pSIC. These numbers are similar
to the picture obtained from the projected DOS analysis.  

\begin{table}
\caption{MLWFs' spreads $\Omega_n$ [in \AA$^2$], and  
contributions to the hole occupations $n_h^{MLWF}$ [in $\%$],
obtained by means of the GGA and the pSIC,
for (Ga,Mn)As in the 64-atoms cell (doping 3$\%$). 
For a comparison, the spreads $\Omega_n$ for the atomic Mn and pure GaAs
are also given.
Numbers of symmetry-equivalent $sp$-lobes are in parenthesis. 
Spin channels are denoted by up- and down-arrows.}
\begin{tabular}{lcc}
\hline
\hline \\[-0.2cm]
 functions  &  GGA  & pSIC  \\[0.1cm]
 \hline \\[0.1cm]
 \multicolumn{3}{c}{Spreads $\Omega_n$} \\[0.2cm]
 Mn: $d_{z^2}$, $d_{x^2-y^2}$ ($\uparrow$) & 0.77$^{(2)}$ & 0.69$^{(2)}$  \\
 Mn: $d_{xy}$, $d_{xz}$, $d_{yz}$ ($\uparrow$) & 
                                   1.20$^{(3)}$ & 1.01$^{(3)}$ \\[0.1cm]
 Mn atom: d$^{5}$ ($\uparrow$)    &  0.46$^{(5)}$  &  0.46$^{(5)}$  \\[0.1cm]
 Mn-As-Ga: $sp^3$ ($\uparrow$)    & 2.82$^{(1)}$ 3.54$^{(3)}$  &  
                     2.95$^{(1)}$ 3.75$^{(3)}$  \\   
 Mn-As-Ga: $sp^3$ ($\downarrow$)  & 2.48$^{(1)}$ 3.30$^{(3)}$  &   
           2.74$^{(1)}$ 3.42$^{(3)}$    \\[0.1cm]
 As-Ga (central): $sp^3$ ($\uparrow$)    
   & 3.35$^{(1)}$ 3.29$^{(3)}$   & 3.64$^{(1)}$ 3.58$^{(3)}$  \\
 As-Ga (central): $sp^3$ ($\downarrow$)  
   & 3.36$^{(1)}$ 3.30$^{(3)}$  &  
   3.64$^{(1)}$ 3.58$^{(3)}$    \\[0.1cm] 
 pure GaAs: $sp^3$      & 3.15$^{(4)}$  &  3.59$^{(4)}$  \\[0.2cm]  
 \multicolumn{3}{c}{Hole occupations $n_h^{MLWF}$} \\[0.2cm]
 Mn: $d_{z^2}$, $d_{x^2-y^2}$ ($\uparrow$) & 0.0$^{(2)}$ & 0.0$^{(2)}$ \\
 Mn: $d_{xy}$, $d_{xz}$, $d_{yz}$ ($\uparrow$) & 4.59$^{(3)}$ & 1.39$^{(3)}$ \\
 Mn-As-Ga: $sp^3$ ($\uparrow$)    & 3.70$^{(1)}$ 2.22$^{(3)}$ & 
                                    1.70$^{(1)}$ 1.67$^{(3)}$  \\
 As-Ga (central): $sp^3$ ($\uparrow$)  & 0.37$^{(1)}$ 0.39$^{(3)}$ &  
                                        0.52$^{(1)}$ 0.56$^{(3)}$  \\[0.1cm]
\hline
\hline
\end{tabular}
\label{spreads}
\end{table}

It is significant that after taking a closer look at the hole bandwidth, 
i.e.  $[{\varepsilon_F},{\varepsilon_{N(\varepsilon)=0}}]$,
for different dopings, we find that in every considered case the system is not strictly insulating.
For 3$\%$ of Mn, the hole bandwidth is 0.3 eV (GGA) or 0.4 eV (pSIC), 
and for 1$\%$ of Mn, it amounts to 0.18 eV (GGA) or 0.28 eV (pSIC). 
The hole bandwidth reduction with decreasing doping is due to increasing
number of states in larger supercells while the number of holes is kept
the same (namely one) - this is independent on whether the 3$d$-states of Mn
reside at the Fermi level or not. We remind, that the maximum of the DOS
of the 3$d$-Mn levels is located 3-4 eV below the Fermi level.

\begin{figure}
\includegraphics[scale=0.25,angle=0.0]{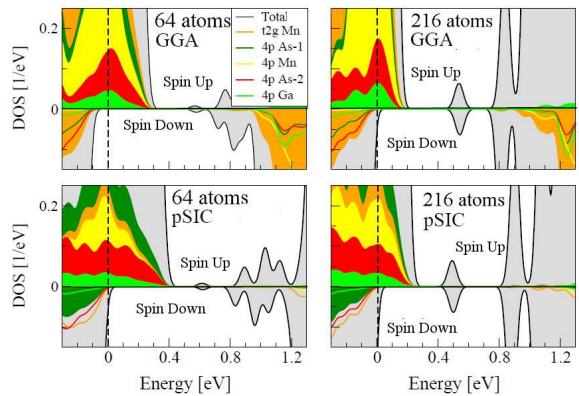}
\caption{(Color online)
Total density of states (DOS) close to the Fermi level
(marked by the vertical dashed-line) and within the energy gap of (Ga,Mn)As,
obtained with the GGA and the pSIC, at 3$\%$ and 1$\%$ dopings.}
\label{gap}
\end{figure}

From ab-initio calculations, we do not observe
separation of the DOS from the top of the valence bands, which was
suggested in few previous works \cite{Samarth,Zunger-new}. 
For very low doping,  
the impurity band within the gap originates from the conduction states.
Detailed picture of the gap region is included in Figure~\ref{gap}.  
Composition of the impurity band within the gap, for the 216-atoms cell and
for both the GGA and the pSIC, consists of the 4$s$-functions 
of Mn, As and Ga atoms, and there is no $d$- or $p$-type add.
Anyway, due to their small bandwidths, these impurity states can 
act as traps for electrons from the Mn interstitials, which are known to be
donors \cite{Erwin,Masek}, 
or these bands can be populated via absorption of photons.

Experimentally, some of diluted samples are conducting and a few among
all samples are insulating; 
the conducting samples have extended HDF and
the insulating samples are characterized by the localized HDF
\cite{Molenkamp}. 
This matches perfectly with our Figure 4 and donor scenario of Ref. \cite{Erwin,Masek}:
\begin{enumerate}
	\item If there is a low concentration of interstitials then holes are not 
compensated, Fermi level lays within the valence band, and samples are 
conducting. 
\item  If there are many donors, then the holes within the valence
band are compensated,  the unpolarized states of the $s$-type  
within the gap (see Figure~\ref{gap} in 216-atoms cases) 
become partially or totally occupied, 
the Fermi level cuts through these states, or between them and 
the valence band, or between the gap states and the conduction band, and   
the samples are insulating.
\end{enumerate}

\section{Summary}

We have used two approaches: the pSIC and the MLWF, in order to analyse in
detail the Fermi level and the gap regions of the (Ga,Mn)As density of states,
and the hole  localization and its chemical character. 

It is demonstrated that the HDF is very delocalized,
especially within the self-interaction corrected scheme.
In diluted case, the HDFs are mainly spread over the crystal volume 
among the impurities, 
not close to the impurities regions. Only at higher concentrations
the HDF resides mainly at the Mn-As complex, 
and this part of the HDF   
has mainly $sp^3$-character centred on the As-neighbours of Mn. 

For very low dopings, the hole states (low energy unoccupied bands) 
are still merged with the valence band,
and the impurity band within the gap forms via separation
from the conduction states. These gap states are unpolarized and 
purely of the $s$-type. Our new result requires experimental confirmation. 

Fermi level is pinned within the valence band for donor-free
samples, or within the localized gap states of $s$-type for donor-rich
samples and low Mn dopings - these samples are insulating and nonmagnetic.
With the above findings, models for Curie temperature in 
dilute magnetic semiconductor (Ga,Mn)As,
which assume an extended hole density function over the valence 
band \cite{Dietl-Tc,Jungwirth,Molenkamp}, are clearly justified.

\section{Acknowledgments}
We would like to thank Prof. R.~R.~Ga\l{}\c{a}zka for encouraging discussion.
Arek Niegowski is kindly acknowledged
for assistance with the computing-system.  
Calculations have been performed in the Interdisciplinary Centre of
Mathematical and Computer Modelling (ICM) of the University of Warsaw
within the grants G47-5 and G47-7 and in Polish Infrastructure of
Informatic Support for Science in European Scientific Space (PL-Grid)
within the projects No. POIG.02.03.00-00-028/08-00 and No. MRPO.01.02.00-12-479/02.


\begin{thebibliography}{99}

\bibitem{RRG} R.~R.~Ga{\l}\c{a}zka and J.~Kossut, {\it
Lecture Notes in Physics} (Springer, Berlin) {133} 1980, p. 245.
\bibitem{Furdyna} J.~K.~Furdyna, J.~Appl.~Phys. {64}(4) (1988) R29;
J.~K.~Furdyna, {\it Semimagnetic Semiconductors and
Diluted Magnetic Semiconductors}, Eds. M.~Averous and M.~Balkanski
(New York: Plenum) 1991.
\bibitem{Ohno} H.~Ohno, Science  {281} (1998) 951.
\bibitem{Dietl-Nature} T.~Dietl, {Nature~Materials} {9} (2010) 965.
\bibitem{Sato-RMP}
K.~Sato, L.~Bergqvist, J.~Kudrnovsk\'y, P.~H.~Dederichs,
O.~Eriksson, I.~Turek, B.~Sanyal, G.~Bouzerar, H.~Katayama-Yoshida,
V.~A.~Dinh, T.~Fukushima, H.~Kizaki, R.~Zeller, {Rev.~Mod.~Phys.} {82}
(2010) 1633.
\bibitem{Dietl-Science}
T.~Dietl, H.~Ohno, F.~Matsukura, J.~Cibert, and D.~Ferrand,
{Science} {287} (2000) 1019.
\bibitem{Dietl-Tc}
T.~Dietl, H.~Ohno, F.~Matsukura, {Phys.~Rev.~B} {63} (2001) {195205}.
\bibitem{RMP-TD}
T.~Dietl, H.~Ohno, {arXiv:1307.3429} (2013).
\bibitem{oiwa}
A.~Oiwa, S.~Katsumoto, A.~Endo, M.~Hirasawa, Y.~Iye,
 F.~Matsukura, A.~Shen, Y.~Sugawara, H.~Ohno,
{Phys.~B:~Cond.~Matt.} {249<96>251} (1998) {775}.
\bibitem{hamaya}
K.~Hamaya, T.~Taniyama, T.~Koike, Y.~Yamazaki,
{J.~App.~Phys.} {99} (2006) {123901}.
\bibitem{mac}
F.~Maccherozzi, G.~Panaccione, G.~Rossi, M.~Hochstrasser,
 M.~Sper, M.~Reinwald, G.~Woltersdorf, W.~Wegscheider, C.~H.~Back,
{Surf.~Scien.} {601} (2007) {4283}.
\bibitem{prb}
M.~Schlapps, T.~Lermer, S.~Geissler, D.~Neumaier, J.~Sadowski, D.~Schuh,
W.~Wegscheider, D.~Weiss,
{Phys.~Rev.~B} {80} (2009) {125330}.
\bibitem{reid}
A.~H.~M.~Reid, G.~V.~Astakhov, A.~V.~Kimel, G.~M.~Schott, W.~Ossau, K.~Brunner,
A.~Kirilyuk, L.~W.~Molenkamp,  Th.~Rasing,
{App.~Phys.~Lett.} {97} (2010) {232503}.
\bibitem{pSIC}
A.~Filippetti and V.~Fiorentini, {Eur.~Phys.~J.~B} {71} (2009) {139}.
\bibitem{implement}
M.~Wierzbowska and J.~A.~Majewski, {Phys.~Rev.~B} {84} (2011) {245129}.
\bibitem{MLWF}
N.~Marzari and D.~Vanderbilt, {Phys. Rev. B} {56} (1997) {12847};
N.~Marzari, A.~A.~Mostofi, J.~~R.~Yates, I.~Souza and D.~Vanderbilt,
{Rev.~Mod.~Phys.} {84} (2012) {1419}.
\bibitem{Park}
J.~H.~Park, S.~K.~Kwon and B.~I.~Min, {Physica~B} {281} (2000) {703}.
\bibitem{Sandratskii}
L.~M.~Sandratskii, P.~Bruno, and J.~Kudrnovsk\'y, {Phys. Rev. B} {69} (2004) {195203}.
\bibitem{Dublin}
M.~Wierzbowska, D.~S\'anchez-Portal and S.~Sanvito,
{Phys. Rev. B} {70} (2004) {235209}.
\bibitem{Alessio}
A.~Filippetti, N.~A.~Spaldin and S.~Sanvito,
Chemical Physics, 309 (2004) 59;
A.~Filippetti, N.~A.~Spaldin and S.~Sanvito,
J. Magn. and Magn. Mat. 290–291 (2005) 1391.
\bibitem{Szotek}
T.~C.~Schulthess, W.~M.~Temmerman, Z.~Szotek, W.~H.~Butler and
G.~M.~Stocks, {Nature~Materials} 4 (2005) 838.
\bibitem{Hill}
S.~Sanvito, P.~Ordej\'on, N.~A.~Hill, {Phys. Rev. B}, 63 (2001) 165206.
\bibitem{Stroppa}
A.~Stroppa, G.~Kresse, A.~Continenza, {Phys. Rev. B}, 83 (2011) 085201.
\bibitem{Zunger-old}
P.~Mahadevan and A.~Zunger, {Phys.~Rev.~B} {69} (2004) {115211}.
\bibitem{Jungwirth}
T.~Jungwirth, J.~K\"onig, J.~Sinova, J.~Kucera, and A.~H.~MacDonald,
{Phys. Rev. B} {66} (2002) {012402}.
\bibitem{Molenkamp}
T.~Jungwirth, J.~Sinova, A.~H.~MacDonald, B.~L.~Gallagher, V.~Novak,
K.~W.~Edmonds, A.~W.~Rushforth, R.~P.~Campion, C.~Foxon, L.~Eaves, E.~Olejnik,
J.~Masek, S.-R.~Eric~Yang, J.~Wunderlich, C.~Gould, L.~W.~Molenkamp, T.~Dietl
and H.~Ohno, {Phys. Rev. B} {76} (2007) {125206}.
\bibitem{Samarth}
N. Samarth, Nature Mater. {11} (2012) {360}.
\bibitem{Zunger-new}
V.~Fleurov, K.~Kikoin and A.~Zunger, {arXiv:1208.2811} (2012).
\bibitem{Kikoin}
P.~M.~Krstajic, V.~A.~Ivanov, F.~M.~Peeters, V.~Fleurov and K.~Kikoin,
{Europhys. Lett.}, 61 (2003) 235.
\bibitem{DFT}
P.~Hohenberg and W.~Kohn, {Phys.~Rev.} {136} (1964) {B864};
W.~Kohn and L.~J.~Sham, {Phys.~Rev.} {140} (1965) {A1133}.
\bibitem{qe}
P.~Giannozzi et al., {J.~Phys.~Condens.~Matter} {21} (2009) {395502}.
\bibitem{PBE}
J.~P.~Perdew, K.~Burke, M.~Ernzerhof, {Phys. Rev. Lett.} {77} (1996) {3865};
{\em Ibid., p} {78} (1997) {1396}.
\bibitem{USPP}
D.~Vanderbilt, {Phys.~Rev.~B} {41} (1990) {R7892}.
\bibitem{grid}
H.~D.~Monkhorst and J.~D.~Pack, {Phys.~Rev.~B} {13} (1976) {5188}. 
\bibitem{smearing}
N.~D.~Mermin, {Phys.~Rev.} {137} (1965) {A1441};
M.~J.~Gillan, {J.~Phys.~Condens.~Matter} {1} (1989) {689}.
\bibitem{wannier} 
A.~A.~Mostofi, J.~R.~Yates, Y.-S.~Lee, I.~Souza, D.~Vanderbilt, N.~Marzari,
Comput. Phys. Commun. {178} (2008) {685}; {www.wannier.org}.
\bibitem{Arash}
L.~Andrinopoulos, N.~D.~M.~Hine and A.~A.~Mostofi,
{J. Chem. Phys.} {135} (2011) {154105}.
\bibitem{lowdin}
A.~Szabo and N.~S.~Ostlund, {Modern Quantum Chemistry. Introduction to Advanced Electronic
Structure Theory.}, Ed. {Dover Publications INC., Mineola, New York} 1996.
\bibitem{xcrys}
A.~Kokalj, {J.~Mol. Graphics Modelling}, 17 (1999) 176;
A.~Kokalj, {Comp.~Mater.~Sci.}, 28 (2003) 155.
Code available from http://www.xcrysden.org/.
\bibitem{Erwin}
S.~C.~Erwin and A.~G.~Petukhov, {Phys.~Rev.~Lett.}, 89 (2002) 227201.
\bibitem{Masek}
J.~Masek, J.~Kudrnovsk\'y, F.~M\'aca, J.~Sinova, A.~H.~MacDonald,
R.~P.~Campion, B.~L.~Gallagher and T.~Jungwirth,
{Phys. Rev. B} {75} (2007) {045202}.
\end{thebibliography}
\end{document}